\documentclass[twocolumn,showpacs,prl]{revtex4}
\usepackage{graphicx}

\begin{document}

\title{Interaction-free measurements with superconducting qubits}

\author{G.~S.~Paraoanu}\email{paraoanu@phys.jyu.fi}

\affiliation{NanoScience Center and Department of Physics, University of Jyv\"askyl\"a,
P.O.~Box 35 (YFL), FIN-40014 University of Jyv\"askyl\"a, Finland} 

\affiliation{Institute for Quantum Optics and Quantum Information, ICT-Geb\"aude, Technikerstra\ss e 21a, A-6020 %%@
Innsbruck, Austria}

\begin{abstract}

An interaction-free measurement protocol is described for a quantum circuit 
consisting of a superconducting qubit and a read-out Josephson junction.
By measuring the state of the qubit one can ascertain the presence of
a current pulse through the circuit at a previous time without any 
energy exchange between the qubit and 
the pulse.

\end{abstract}


\maketitle

%\section{Introduction}

We know from classical optics that when an object blocks one arm of an interferometer the fringes disappear. In %%@
quantum physics, this effect acquires far more subtle features, and, counter-intuitively, it does not originate %%@
from the unavoidable disturbance on the photon trajectory or random phases due to scattering, but from the %%@
possibility of obtaining which-path information. Indeed, in the quantum treatment the interference would disappear %%@
if one can in principle obtain information about which path the light went through, even if this information is not %%@
actually extracted \cite{quantumoptics}. Quantum processes therefore result not only in changes in observables such %%@
as position and momentum (due to energy exchange), but, more importantly, in establishing new correlations between
parts of the system. Recently, a lot of work has been put into harnessing the power of these correlations for %%@
performing computational tasks
which are very difficult to implement on classical computers. Superconducting qubits based on the Josephson effect %%@
have been
proposed \cite{reviews} as the elements of future quantum computers, based on macroscopic quantum coherence effects %%@
in charge and flux devices \cite{coherence}. Several species of superconducting qubits are currently
under study, for example charge qubits \cite{charge}, phase qubits \cite{phase}, flux qubits \cite{flux}, and a %%@
mixed charge-flux version called Quantronium \cite{vion}. This last type has a very large decoherence time (more %%@
than 500ns), and it will be the main focus of this paper. 
Besides quantum computing, fundamental research such as testing quantum mechanics at the macroscopic level  
is an important direction envisioned decades ago \cite{leggett}, with progress in this direction now enjoying a %%@
firm experimental basis. 

\begin{figure}[htb]
\includegraphics[width=75truemm]{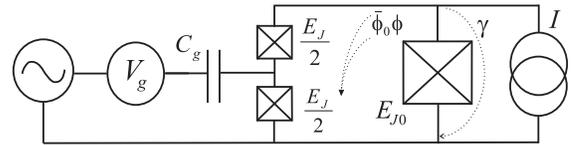}
\caption{Quantronium circuit with flux-current compensation.}
\label{schematicifm}
\end{figure}

In this paper we propose an experiment in which a Quantronium device (Fig. \ref{schematicifm}) could be used to %%@
ascertain the presence of a small pulse of electric current without any disturbance due to energy exchange with the %%@
continuum of states outside the washboard potential well in which the qubit
is localized. An experiment of this type is feasible with the current Quantronium setup, and it would constitute a %%@
test, at the 
macroscopic level, of a strongly nonclassical prediction of quantum mechanics.
The proposal is based on the 
interaction-free measurement scheme proposed by Elitzur and Vaidman for 
optical Mach-Zehnder interferometers \cite{elitzurvaidman}. This realization has found interesting applications in %%@
interaction-free imaging of objects 
\cite{imaging}, and more recently in quantum computing \cite{comp}.

\begin{figure}[htb]
\includegraphics[width=75truemm]{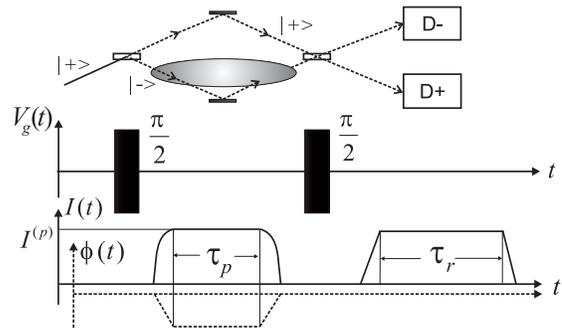}
\caption{Mach-Zehnder interferometric setup (upper schematic) and the Quantronium pulse sequence (graphs below) 
with the axes of the flux graph slightly displaced for clarity.}
\label{pulsesifm}
\end{figure}

In the optical setup (upper schematic in Fig. \ref{pulsesifm}), a balanced Mach-Zehnder interferometer is %%@
constructed from two 50\% beam splitters, two perfectly reflecting mirrors, and two detectors $D+$ and $D-$ ($+$ %%@
and $-$ are the directions corresponding respectively to paths along the upper and lower arms of the %%@
interferometer). In the absence of an object in the lower arm this arrangement produces a destructive interference %%@
at the detector $D+$ (the state of the photon, which is initially $|+>$, becomes $(|+> + i|->)/\sqrt{2}$ inside the %%@
interferometer and $i|->$ after the second beam splitter). If a quantum ultrasensitive object ({\it i.e.} triggered %%@
by the absorption of a single photon) is present, in 50\% of the cases there will be no absorption event: the %%@
photon which has traveled in the upper arm of the interferometer
must be then in the state $|+>$ (it did not "interact" with the object), and it will emerge from the 
second beam splitter in the state $(|+> + i |->)/\sqrt{2}$,
having now a 50\% chance of being detected by $D+$ \cite{elitzurvaidman}. As a result, when $D+$ is triggered we %%@
can be certain about the presence of an object in a region in space without having to exchange energy ({\it e.g.} %%@
by photon absorption) with that object. The success rate (the fraction of photons detected by $D+$) is 25\%.

A Quantronium device (Fig. \ref{schematicifm}) consists of a split Cooper pair box of total capacitance %%@
$C_{\Sigma}$ and Josephson energy $E_{J}/2$ per junction, operated at a
gate voltage $V_g$ and excited by microwave radiation coupled through the gate capacitance $C_g$. The box is %%@
connected in parallel with a read-out Josephson junction $E_{J0}$ of capacitance $C_{0}$. 
The Hamiltonian of the circuit is \cite{vion}
\begin{eqnarray}
H &=& \frac{1}{2C_{\Sigma}}(q - C_{g}V_{g})^{2} - E_{J}\cos\frac{\gamma + \phi}{2}\cos\theta \nonumber \\ && + %%@
\frac{Q^2}{2C_{0}}-
E_{J0}\cos\gamma -I\bar{\phi}_{0}\gamma ,\label{hhh}
\end{eqnarray}
where $Q$ and $\gamma$ are the charge and the phase across  
the large junction, $I$ is the bias current, $\phi\bar{\phi}_{0}$ is the externally applied magnetic flux through %%@
the loop, $\bar{\phi}_{0}=\hbar /2 e$ is the reduced flux quanta,
and \{$q$,$\theta$\} is the pair of conjugate variables (commutation $[\bar{\phi}_{0}\theta ,q] = i\hbar$) %%@
corresponding to the split Cooper pair box. The read-out junction is in the large-capacitance regime, $C_{0}\gg %%@
e^{2}E_{J0}^{-1}$, where
$Q$  becomes a continuous operator $Q=-2ei\partial /\partial \gamma$.
A two-level system (the qubit) is realized \cite{reviews,charge} at the
gate voltage $V_{g}=e/C_{g}$, where the charging-energy degeneracy of
the eigenvectors $|0>$ and $|1>$ of the operator $q$ ($q|0> = 0$, $q|1> = 2e|1>$)
is lifted by the tunneling 
term $- E_{J}\cos (\gamma /2 + \phi /2)\cos\theta$. The eigenvectors of this tunneling term (the qubit levels)
are 
denoted by $|+>$ (ground state) and $|->$ (excited state), $|\pm > = (|0>\pm |1>)/\sqrt{2}$, and in this basis the %%@
Hamiltonian Eq. (\ref{hhh}) is of Stern-Gerlach type 
\begin{equation}
H = -\frac{2e^{2}}{C_{0}}\frac{\partial^{2}}{\partial \gamma^{2}} - E_{J0}\cos\gamma -I\bar{\phi}_{0}\gamma - %%@
\frac{E_{J}}{2}\cos\frac{\gamma + \phi}{2}\sigma_{z}.
\end{equation}
The bias current and the flux are externally controlled parameters ($\phi (t)$, $I (t)$) which are manipulated %%@
adiabatically compared to the time-scale of the qubit frequency. 
The eigenvalue-eigenfunction  problem $H \psi_{\pm}(\gamma ) |\pm >= \epsilon_{\pm}\psi_{\pm}(\gamma )|\pm >$ can %%@
be thus solved at every moment $t$, where, to keep the notation simple, we will specify the corresponding values of %%@
($\phi$, $I$) each time we refer to $\epsilon_{\pm}$ or to the instantaneous eigenvector $\psi_{\pm} (\gamma)$. For %%@
example, at $I=\phi = 0$ the states $\psi_{\pm}(\gamma)$ do not differ much from one another, the phase $\gamma$ is %%@
almost classical-like ($\gamma \approx 0$), and the qubit energy is $\epsilon = \epsilon_{-}-\epsilon_{+} \approx %%@
E_{J}$ \cite{vion}. Adiabatic excursions at nonzero values of $I$ help differentiate between the states $\pm$: a %%@
typical measurement protocol for this circuit proceeds by raising the bias current 
to a value close to the critical current $E_{J0}\bar{\phi}_{0}^{-1}$ of the large read-out junction for a certain %%@
read-out time $\tau_{r}$. The junction then can tunnel in the running-wave state \cite{paraoanu} with a switching %%@
rate $\Gamma_{\pm}^{(r)}$ \cite{vion} which depends on the state $|\pm>$ of the qubit .

The idea of the proposed experiment is to create the time-equivalent of a Mach-Zehnder interaction-free experiment %%@
by inserting a bias pulse (referred to in the following as "interaction pulse") inside a Ramsey sequence of $\pi %%@
/2$ microwave gate pulses. Ramsey techniques are common to many fields of physics; they can be interpreted as %%@
two-paths interferometry, the first $\pi /2$ pulse separating the paths and the second rejoining them to test what %%@
has happened in between.
The experiment proceeds as follows (Fig. \ref{pulsesifm}): at all times, the system is kept at the charge %%@
degeneracy point $C_{g}V_{g}=e$ where the decoherence due to gate voltage fluctuations is zero in the first order. %%@
A first $\pi /2$ pulse initializes the qubit in an equal-weight superposition of the 
states $|+>$ and $|->$. 
After that the bias current is increased adiabatically to a value $I^{(p)}$ that allows tunneling during a time %%@
interval $\tau_{p}$, then turned off back to zero.
Finally, another $\pi /2$ pulse is applied and after that a standard switching current measurement sequence %%@
\cite{vion,paraoanu} follows.
Therefore, in this proposal the
role of the ultrasensitive quantum object is played by the quasi-continuum of modes outside the well of the %%@
washboard potential
of the large junction, into which the phase $\gamma$ is allowed to tunnel during the interaction pulse.

The duration of the whole procedure must be less 
than the decoherence time $T_{2}\approx 500$ns, and the value of $I^{(p)}$ has to be such that the dominant %%@
switching process is macroscopic quantum tunneling (and not thermal activation).
It is simple to maintain the qubit at the optimal charge parameter point $C_{g}V_{g}=e$, but changes in the bias %%@
current will displace the system along the phase direction \cite{vion}. We propose to compensate this displacement %%@
by adding a magnetic field pulse such that $\phi (t)= - \gamma_{m}(t)$, where $\gamma_{m}(t)$ is the phase %%@
corresponding to the minimum of the washboard well potential of the large junction at a time t during the current %%@
pulse, defined by $\sin\gamma_{m}(t)=I (t)\bar{\phi}_{0}/E_{J0}$, $0<\gamma_{m}(t)<\pi /2$.
As a result, there will be no linear longitudinal noise component due to fluctuations of the flux in the loop ({\it %%@
e.g.} vortices moving in and out, fluctuations 
of the externally applied magnetic field) and the decoherence time $T_{2}$ is expected to stay of the order of %%@
0.5$\mu$s. Technically, it is easier to apply a magnetic field pulse and then model the current bias pulse 
according to the relation above ($I (t) = - E_{J0}\bar{\phi}_{0}^{-1}\sin\phi(t)$), as shown in Fig. %%@
\ref{pulsesifm}.
This simultaneous manipulation of the bias current and flux results in a different physics than that of the 
usual read-out pulse. In the last case, the two states of the qubit are distinguished by the appearance of small %%@
electrical currents that substract from or add to the externally imposed
bias current. In our procedure, during the excursion toward the switching point, these currents are maintained to %%@
zero to reduce decoherence; however, the two states are still distinguishable by different values of the plasma %%@
oscillation frequency and of the tunneling barrier height, which lead to different macroscopic quantum tunneling %%@
probabilities. Finally, one can notice that
the qubit Larmor frequency $\epsilon$ remains unchanged during the compensated interaction pulses, therefore no 
extra phase differences are introduced in the Ramsey interference pattern.

At the plateau $I^{(p)}$ of the interaction pulse, 
by expanding the potential energy around $\gamma_{m}^{(p)} = \arcsin\left(I^{(p)}\bar{\phi}_{0}/E_{J0}\right)$, we %%@
find 
the Josephson plasma frequency 

and the corresponding washboard potential height
$
\Delta{U}_{\pm}^{(p)} = (2/3)E_{J0}\left[\sqrt{1-\left(I^{(p)}\bar{\phi}_{0}/E_{J0}\right)^{2}}\pm %%@
E_{J}/8E_{J0}\right]^{3}$.
The switching rate can be calculated in the WKB approximation \cite{leggett,paraoanu} and will be  different for %%@
the two qubit states,
$
\Gamma_{\pm }^{(p)} = 52\sqrt{\Delta U_{\pm}^{(p)}/\hbar\omega_{\pm}^{(p)}}(\omega_{\pm}^{(p)}/2\pi %%@
)\exp{\left(-7.2\Delta U_{\pm}^{(p)} /\hbar\omega_{\pm}^{(p)}\right)}$.

We start with the qubit in the ground state $\psi_{+}(\gamma )|+>$; after the first $\pi /2$ Ramsey pulse this %%@
state (all the states from now on will be written in the interaction picture) changes into the superposition %%@
$(\psi_{+}(\gamma )|+> + i\psi_{-}(\gamma ))|->/\sqrt{2}$ \cite{quantumoptics,vion}.
In the absence of an interaction pulse,
one can check that after the second Ramsey $\pi /2$ pulse the state of the system is $i\psi_{-}(\gamma )|->$ %%@
\cite{quantumoptics} (destructive interference toward the $D+$ detector, in the Mach-Zehnder setup). Therefore, %%@
during the read-out pulse of length $\tau_{r}$, the non-switch probability is $\exp (-\Gamma^{(r)}_{-}\tau_{r})$. 

In the case in which we have an interaction pulse, the status of the quantum circuit immediately at the end of
the interaction pulse depends on whether the large junction has switched or not. Immediately after the interaction %%@
pulse, if the large junction did not switch, which happens with probability 
${\cal N} = \left[\exp(-\Gamma_{+}^{(p)}\tau_{p}) + \exp(-\Gamma_{-}^{(p)}\tau_{p})\right]/2$,
the state of the circuit is $(2{\cal N})^{-1/2}\exp(-\Gamma^{(p)}_{+}\tau_{p} /2)\psi_{+}(\gamma )|+> + i(2{\cal %%@
N})^{-1/2}\exp(-\Gamma^{(p)}_{-}\tau_{p} /2)\psi_{-}(\gamma )|->$.
The second Ramsey pulse transforms this state into  
\begin{eqnarray}
(2\sqrt{{\cal N}})^{-1}\left(e^{-\Gamma^{(p)}_{+}\tau_{p} /2}-e^{-\Gamma^{(p)}_{-}\tau_{p} %%@
/2}\right)\psi_{+}(\gamma )|+ > \nonumber \\ 
+(2\sqrt{{\cal N}})^{-1}\left(e^{-\Gamma^{(p)}_{+}\tau_{p} /2}+e^{-\Gamma^{(p)}_{-}\tau_{p} %%@
/2}\right)\psi_{-}(\gamma )|- > .\label{stare}
\end{eqnarray}
Finally, after the read-out pulse, the probability that the system did not switch during the entire sequence of %%@
pulses is a conditional probability obtained by multiplying the probability ${\cal N}$ that the system did not %%@
switch during the first pulse with the nonswitch probabilities during the read-out pulse for the state Eq. %%@
(\ref{stare}), with the result
\begin{eqnarray}
&& \left(e^{-\Gamma^{(p)}_{+}\tau_{p} /2}-e^{-\Gamma^{(p)}_{-}\tau_{p} /2}\right)^{2}e^{-\Gamma^{(r)}_{+}\tau_{r}} %%@
/4 \nonumber \\
 &+& \left(e^{-\Gamma^{(p)}_{+}\tau_{p} /2}+e^{-\Gamma^{(p)}_{-}\tau_{p} %%@
/2}\right)^{2}e^{-\Gamma^{(r)}_{-}\tau_{r}}/4.\label{ioane}
\end{eqnarray}

We analyze now the read-out pulse sequence. To avoid the case in which non-switching events occur in the absence of %%@
the 
interaction bias pulse, we have to set the corresponding probability $\exp (-\Gamma^{(r)}_{-}\tau^{(r)})$
as small as possible: suppose we consider this spurious non-switching rate as satisfactory when it is of the order %%@
of $10^{-3}$.
To determine the read-out bias pulse parameters, we will use directly the experimental data reported so far %%@
\cite{vion}. 
Suppose we use a bias read-out current of 1.11$\mu$A ($94.9\%$ of the critical current %%@
$E_{J0}\bar{\phi}^{-1}_{0}$); then at 14 mK the switching can be estimated as $\Gamma^{(r)}_{-} = 15$MHz, and %%@
$\Gamma^{(r)}_{+} = 2$MHz. For a read-out time $\tau^{(r)} = 460$ns we obtain $\exp (-\Gamma^{(r)}_{-}\tau^{(r)}) = %%@
10^{-3}$ and 
$\exp (-\Gamma^{(r)}_{+}\tau^{(r)}) = 0.4$. 

Let us now focus on the interaction pulses. For an interaction bias pulse of height
$I^{(p)}\bar{\phi}_{0}/E_{J0}= 95.2\%$, we obtain $\Gamma^{(p)}_{+}=2.22$MHz and $\Gamma^{(p)}_{-}=6.58$MHz. Since
$\exp (-\Gamma^{(r)}_{-}\tau^{(r)})$ is negligible, we find from Eq. (\ref{ioane}) that the maximum success rate is %%@
obtained when the difference between the nonswitching probabilities of the two states,  
$\exp (-\Gamma^{(p)}_{+}\tau_{p}/2) - \exp (-\Gamma^{(p)}_{-}\tau_{p}/2)$
is maximal. For $\tau_{p}\approx 250$ns -- a value of the same order as that used routinely in switching current %%@
experiments -- we reach  
$\left[\exp (-\Gamma^{(p)}_{+}\tau_{p}/2) - \exp (-\Gamma^{(p)}_{-}\tau_{p}/2)\right]^{2}/4=2.5\%$. Note that the %%@
pulse duration is below 500 ns; we have checked also that the thermal activation rate is negligible.
Now, taking into account the $\exp (- \Gamma^{(r)}_{+}\tau^{(r)}) = 0.4$ read-out discrimination factor 
estimated above, we obtain as the final result a success rate of $1\%$, much larger than the 
error due to spurious nonswitching events and easily detectable experimentally. In a typical experiment recording %%@
about $10^{4}$ events per second, this corresponds to
100 successful events every second. This success rate is smaller than the theoretical maximum of 25$\%$, resulting %%@
from the imperfect discrimination between the states $|+ >$ and $|- >$. Due to this, 
our estimates show that the success rate cannot be improved by using the quantum Zeno effect \cite{elitzurvaidman}. %%@
In a Mach-Zehnder setup, this would correspond to the quantum ultrasensitive object interacting with
different probabilities with photons propagating in both arms of the interferometer.

It is instructive to consider the case in which a phase error $\Delta = \hbar^{-1}\int \delta\epsilon (t) dt$ %%@
($\delta\epsilon (t)$ is the change of the qubit energy at each instant $t$)
is introduced due to an imperfect 
compensation $\phi (t) \neq -\gamma_m(t)$.  This will produce a dephasing of the qubit with respect to the %%@
microwave radiation at the second $\pi /2$ Ramsey pulse. Immediately after the interaction pulse, the circuit is in %%@
the state
$(2{\cal N})^{-1/2}\exp(i\Delta /2)\exp (-\Gamma^{(p)}_{+}\tau_{p} /2)\psi_{+}(\gamma )|+> + i(2{\cal %%@
N})^{-1/2}\exp(-i\Delta /2)\exp(-\Gamma^{(p)}_{-}\tau_{p} /2)\psi_{-}(\gamma )|->$, and after the second Ramsey %%@
pulse the state assumes the form
\begin{eqnarray}
(2{\sqrt{\cal N}})^{-1}\left(e^{i\Delta /2-\Gamma^{(p)}_{+}\tau_{p}/2}-e^{-i\Delta %%@
/2-\Gamma^{(p)}_{-}\tau_{p}/2}\right)\psi_{+}(\gamma )|+ >\nonumber \\+i(2{\sqrt{\cal N}})^{-1}\left(e^{i\Delta %%@
/2-\Gamma^{(p)}_{+}\tau_{p}/2}+e^{-i\Delta /2 -\Gamma^{(p)}_{-}\tau_{p}/2}\right)\psi_{-}(\gamma )|- >. \nonumber
\end{eqnarray}
The probability of not observing a switch during the entire sequence (with the read-out strategy $\exp %%@
(-\Gamma^{(r)}_{-}\tau^{(r)})\approx 0$ adopted before), 
is obtained as
\begin{eqnarray}
\left(e^{-\Gamma^{(p)}_{-}\tau_{p} /2} 
-e^{-\Gamma^{(p)}_{+}\tau_{p} /2}\right)^{2}e^{-\Gamma_{+}^{(r)}\tau^{(r)}}/4 \nonumber \\ + e^{-(\Gamma^{(p)}_{+} %%@
+ 
\Gamma^{(p)}_{-})\tau_{p} /2}\sin^{2}(\Delta /2)e^{-\Gamma_{+}^{(r)}\tau^{(r)}} . \label{fain}
\end{eqnarray}
One can now clearly distinguish between classical interference (oscillations of probability as a sine function, the %%@
second term of the expression in square brackets above) and the pure interaction-free effect (the first term in the %%@
sum above). 
Classical optical interferometry allows the detection of a transparent object due to its index of refraction
(resulting in different optical path lengths). Interaction-free detection allow us to assert the presence of %%@
perfectly opaque objects without any photon being absorbed. In general, if $\Delta$ is nonzero, both effects are %%@
present,
as shown in Eq. (\ref{fain}). In the case of the parameters suggested above for the interaction pulse, a dominant %%@
interaction-free effect is obtained if $\sin^{2}(\Delta /2) \ll 7.5\%$ (within the reach of present-day experiments %%@
\cite{vion}). The experimental setup can be also calibrated first with a dummy interaction pulse with the same %%@
parameters as the real one, except for the value of the plateau current which should be slightly lower than the one %%@
of the real interaction pulses $I^{(p)}$, such that no switching occurs in the interval $\tau_{p}$. Thus, the %%@
experimentalist can use this pulse to adjust the shape of the external flux pulse so that only switching events are %%@
recorded.
This is helped as well by the fact that the changes $\delta\epsilon (t)$ are second-order in the error representing %%@
the mismatch between the 
desired current pulse shape $I(t)= - E_{J0}\bar{\phi}_{0}^{-1}\sin\phi(t)$ and the bias current signal that %%@
effectively reaches the 
sample, a direct consequence of the very definition of the optimal working point at which we are attempting to %%@
operate (first order changes in the qubit energy due to phase errors and fluctuations are zero). Also, the errors %%@
$\delta\epsilon (t)$ occur mostly during the relatively short raise and fall times (50 ns) and much less at the %%@
plateau - where the flux and the current being constant it is easier to ensure precisely the compensation. 

Finally, one can realize interaction-free experiments with even less control over the accumulated phase difference %%@
$\Delta$, using interaction pulses with slightly higher $I^{(p)}$ which tend to nullify predominantly the %%@
interferometric term. The price to pay for this is a reduction in the success rate. As an example, if %%@
$I^{(p)}\bar{\phi}_{0}/E_{J0} = 96\%$ and for a maximum error $\Delta = \pm \pi$, the interference term becomes 7 %%@
times smaller than the interaction-free term (now also reduced to 0.27\%).

In  conclusion, two fundamental physical processes, interferometry and tunneling, can be combined to demonstrate %%@
the equivalent of the non-classical interaction-free detection scheme for a superconducting quantum circuit. The %%@
crossover between standard interference effects and the interaction-free phenomenon is also discussed.

%\section{Sample fabrication}

%\section{Conclusion}

%\section{Acknowledgements}  
%\acknowledgements
  
This work was supported by the Academy of Finland (Acad. Res. Fellowship no. 00857 and projects no. 7111994 and %%@
7205476) and EU SQUBIT-2 (IST-1999-10673).
The author is grateful to P.~Zoller for supporting a 
research visit at IQOQI Innsbruck.


\begin{thebibliography}{99}




\bibitem{quantumoptics} M.~O.~Scully and M.~Suhail Zubairy, {\it Quantum Optics} (Cambridge University Press, %%@
Cambridge 1997).

\bibitem{reviews} Y.~Makhlin, G.~Sch$\ddot{\mathrm{o}}$n, and A.~Shnirman,
 Nature {\bf 386}, 305 (1999); J.~E.~Mooij {\it et. al.}, Science {\bf 285}, 1036 (1999); Yu.~Makhlin, %%@
G.~Sch$\ddot{\mathrm{o}}$n, and A.~Shnirman,
Rev.~Mod.~Phys. {\bf 73}, 357 (2001).


\bibitem{coherence}Y.~Nakamura, Y.~A.~Pashkin, and J.~S.~Tsai, 
 Nature, {\bf 398}, 786 (1999); 
  C.~H.~van~der~Wal {\it et. al.}, Science {\bf 290}, 773 (2000);
 J.~R.~Friedman {\it et. al.}, Nature {\bf 406},
 43 (2000).
 
\bibitem{charge} Yu.~A.~Paskin {\it et. al.}, Nature {\bf 421}, 823 (2003); T.~Yamamoto {\it et. al.}, Nature {\bf %%@
425}, 941 (2003); T.~Duty, D.~Gunnarsson, K.~Bladh, and P.~Delsing, Phys. Rev. B {\bf 69}, 140503(R) (2004).

\bibitem{phase}J.~M.~Martinis, S.~Nam, J.~Aumentado, and C.~Urbina, Phys.~Rev.~Lett. {\bf 89}, 117901 (2002).

\bibitem{flux}I.~Chiorescu, Y.~Nakamura, 
 C.~J.~P.~M.~Harmans, and J.~E.~Mooij, Science {\bf 299}, 1869 (2003).

\bibitem{vion} D.~Vion {\it at. al}, Science {\bf 296}, 886 (2002);  A.~Cottet {\it et. al.}, Physica C {\bf 367}, %%@
197 (2002); D.~Vion {\it et. al.} Phys. Scr. T {\bf 102}, 162 (2002); A. Cottet, Ph. D. thesis, Univ. Paris VI %%@
(2002); E.~Collin {\it et. al.}, Phys. Rev. Lett. {\bf 93}, 157005 (2004);
 G.~Ithier, {\it et. al.} Phys. Rev. B {\bf 72}, 134519 (2005).

\bibitem{leggett} A.~Widom, J. Low Temp. Phys. {\bf 37}, 449 (1979); A.~J.~Leggett and A.~Garg, Phys.~Rev.~Lett. %%@
{\bf 54}, 857 (1985); A.~J.~Leggett, J.~Phys.: Condens. Matter
{\bf 14}, R415 (2002).
 
\bibitem{elitzurvaidman} A.~Elitzur and L.~Vaidman, Found.~Phys. {\bf 23}, 987 (1993); L.~Vaidman, Found.~Phys. %%@
{\bf 33}, 491 (2003); L.~Vaidman, Quantum Opt. {\bf 6}, 119 (1994); P.~Kwiat {\it et. al.}, Phys.~Rev.~Lett. {\bf %%@
74}, 4763 (1995); P.~G. ~Kwiat {\it et. al.} Phys.~Rev.~Lett. {\bf 83}, 4725 (1999). 

\bibitem{imaging} A.~G.~White, J.~A.~Mitchell, O.~Nairz, and P.~G.~Kwiat, Phys.~Rev.~A {\bf 58}, 605 (1998); J.-S. %%@
Jang,
Phys.~Rev.~A {\bf 59} 2322 (1999);
G.~Mitchison and S.~Massar, Phys.~Rev.~A {\bf 63}, 032105 (2001).

\bibitem{comp} O. Hosten {\it et. al.} Nature {\bf 439}, 949 (2006).  

\bibitem{paraoanu} J.~Clarke {\it et. al.}
Science {\bf 239}, 992 (1988); G.~S.~Paraoanu,
Phys.~Rev.~B {\bf 72}, 134528 (2005).



\end{thebibliography}
\end{document}